\documentclass[traditabstract]{aa} 
                                   
\usepackage{graphicx}
\usepackage{multirow}
\usepackage{txfonts}
\usepackage{natbib}
\bibpunct{(}{)}{;}{a}{}{,} 

\begin{document}
   \title{Radiative forces on macroscopic porous bodies in protoplanetary disks: laboratory experiments}

   \author{Christoph Duermann \and Gerhard Wurm \and Markus Kuepper}

   \institute{Fakulty of Physics, University of Duisburg-Essen,
              Lotharstr. 1, D-47057 Duisburg \\
              \email{gerhard.wurm@uni-due.de}
             }

   \date{Received ; accepted }
 
\abstract{
In optically thin parts of protoplanetary disks photophoresis is a significant force not just for dust grains, 
but also for 
macroscopic bodies. The absolute strength on the supposedly highly porous objects is not known in detail as yet. We set up a low pressure torsion balance and studied photophoretic forces down to 100\,nN on plates at a light flux of $100\,\rm W/m^2$. We investigated the dependence on plate dimensions and on ambient pressure and considered the influence of channels through the plates. As samples for full (no channel) plates we used tissue with 2\,mm thickness and circular shape with diameters of 10\,mm, 30\,mm and 50\,mm. The influence of channels was probed on rectangular-shaped circuit boards of 35\,mm\,$\times$\,35\,mm area and 1.5\,mm thickness. The number of channels was 169 and 352. The pressure was varied over three decades between 0.001 and 1\,mbar.  At low pressure, the absolute photophoretic force is proportional to the cross section of the plates. At high pressure, gas flow through the channels enhances the photophoretic force. The pressure dependence of the radiative force can (formally) be calculated by photophoresis on particles with a characteristic length. We derived two characteristic length scales $l$ depending on the plate radius $r_1$, the channel radius $r_2$, and the thickness of the plate which equals the length of the channel $d$ as $l=r^{0.35}\cdot d^{0.65}$. The highest force is found at a pressure $p_{max} = 15 \cdot l^{-1}$ Pa mm. In total, the photophoretic force on a plate with channels can be well described by a superposition of the two components: photophoresis due to the overall size and cross section of the plate and photophoresis due to the channels, both with their characteristic pressure dependencies. We applied 
these results to the transport of large solids in protoplanetary disks and found that the influence of porosity on the photophoretic force can reverse the inward drift of large solids, for instance meter-sized bodies, and push them outward within the optically thin parts of the disk.}

\titlerunning{Photophoretic Force on Channel Plates}
\maketitle

\section{Introduction}
For the past several years photophoretic forces have been considered to
influence particle transport in protoplanetary disks \citep{krauss2005, wurm2006,herrmann2007,  krauss2007, mousis2007,  takeuchi2008, wurm2009, wurm2009b, moudens2011, vonBorstel2012, debeule2013}.
Whenever particles embedded in a gaseous environment are illuminated and absorb some of the radiation, the temperature gradient along the particle and the interaction with gas molecules leads to a force that -- in most cases -- is directed away from the light source. A particle motion induced by  such radiometric forces is commonly called photophoresis \citep{ehrenhaft1918, rohatschek1995}. 

The strength of photophoresis on cm-size particles is easily demonstrated by a Crookes radiometer \citep{crookes1874} and can be a factor of 100 higher than stellar gravity for 1 Pa pressure \citep{krauss2005}. Supported by photophoresis, large bodies drifting inward would slow down or even reverse motion close to the inner edge
of a disk and move outward as shown below. 

Unfortunately, the complexity of photophoresis increases with the complexity of particle shapes. Large objects in protoplanetary disks for instance are highly porous \citep{Meisner2012}. Collisions of larger aggregates might also lead to macroporosity of large objects  \citep{Kothe2013}. The question therefore is how such large complex bodies react when they are subjected to photophoresis. Is an individual constituent of an aggregate important, or does the whole aggregate set the radiative force at a given ambient gas pressure? Do large pores or channels through an aggregate increase or decrease the photophoretic force, and if so, in which pressure range?

\begin{figure}
	\centering
	\includegraphics[width=0.4\textwidth]{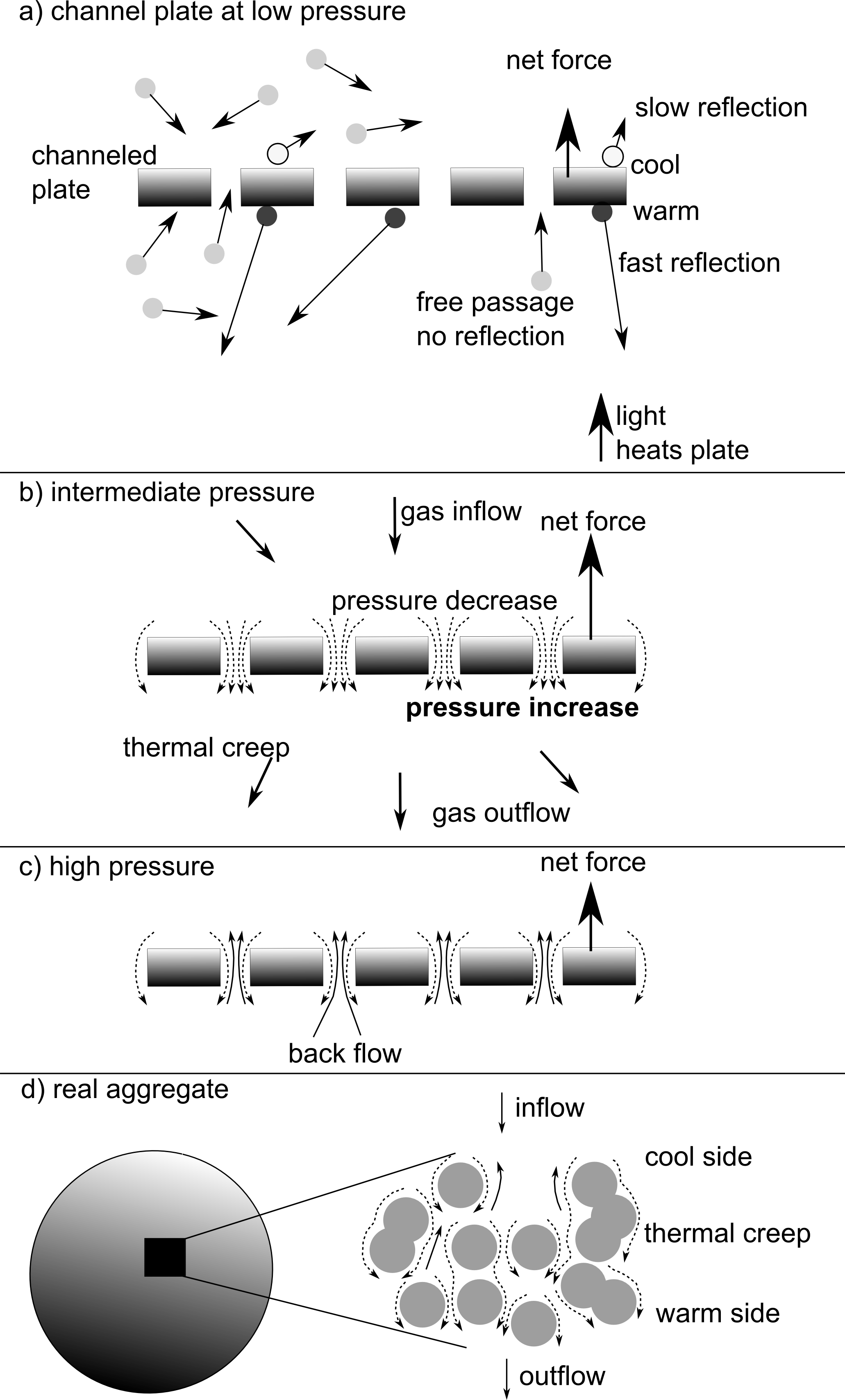}
	\caption{Idea of photophoretic force enhancement by channels and macro pores
	a) low pressure: only the real cross-sectional area counts, individual molecule
	collisions and momentum transfer; b) intermediate pressure: thermal creep in channels leads to increase of pressure at warm side; c) high pressure: thermal creep only acts close to the inner channel walls; pressure increase can be balanced more easily by back flow through the center part of the channel; d) thermal creep in more realistic porous bodies in protoplanetary disks (dust aggregates, chondrule aggregates, larger rubble piles).}
	\label{fig:idea}
\end{figure}

Temperature gradients across a particle can lead to thermal creep (gas flow) through channels \citep{Alexeenko2006, vargo1999, knudsen09, muntz2002, Han2007}. This increases the photophoretic force on the particle in certain pressure ranges that have been studied for instance by \citet{Kuepper2013} under microgravity or in this work.
This paper is a first attempt at quantifying the influence of channels
of a given size through a larger body on the photophoretic force in a systematic way. 
The idea of photophoresis for large (macro)-porous bodies is sketched in fig. \ref{fig:idea} {(see text below for
details of the specific regimes sketched)}.

To quantify this we set up a torsion balance to measure the photophoretic force
on cm-size plates with and without channels and characterized the pressure dependence of the photophoretic force. In a first application we show the significance of pores for transport in protoplanetary disks.

\section{Photophoresis}

Photophoresis for spherical particles has been considered in detail for the whole range of ambient pressure from the free molecular flow regime at low pressure to hydrodynamic flow at high pressure \citep{Beresnev1993, rohatschek1995}. Spherical particles have the advantage that only one size, the particle radius $a$ enters.
A semi-empirical description for photophoresis over the whole pressure range has been worked out by \citet{rohatschek1995} for a perfectly absorbing sphere given as
\begin{equation}
	F_{Ph} = \frac{\left(2 + \delta\right)F_{max}}{\frac{p}{p_{max}} + \delta + \frac{p_{max}}{p}},
	\label{eq:Rohatschek}
\end{equation}
with 
\begin{eqnarray}
	F_{max}  &=  \frac{a^2}{2} D \sqrt{\frac{\alpha}{2}}{\frac{I}{k}},\nonumber \\ 
	p_{max}  &= \frac{3T}{\pi a}D\sqrt{\frac{2}{\alpha}}, \nonumber \\
	D  &=  \frac{\pi \bar{c} \eta}{2T} \sqrt{\frac{\pi \kappa}{3}}\nonumber \\
	\bar{c}  &=  \sqrt{\frac{8R_g T}{\pi \mu}},
	\label{eq:RohatschekParameters}
\end{eqnarray}
where $a$ is the particle radius, $I$ is the radiant flux density, $T$ is the gas temperature, $p$ is the gas pressure, $k$ is the thermal conductivity of the particle, $\alpha$ is the thermal accomodition coefficient, which is often assumed to be 1, $\eta$ is the viscosity of the gas, $\kappa$ is the thermal creep coefficient, which has a value of 1.14 \citep{rohatschek1995}, $R_g$ is the universal gas constant, and $\mu$ is the molar mass of the gas particles. {$\delta$ is a free parameter on the order of $|\delta| \leq 1$ to connect the low-pressure and high-pressure range in different models. For different experimental settings it can be regarded as a free
empirical parameter. However, it mostly influences the width of the transition around $p_{max}$. It does not change the position of the maximum, though.} 

{The transition occuring at $p_{max}$ can also be illustrated in another way. For this we refer to the Knudsen number, which is defined as $\mathit{Kn} = \lambda / a$, where $\lambda$ is the mean free path of the gas molecules.  It ist not immediately evident in eq. \ref{eq:RohatschekParameters}, but the pressure
 $p_{max}$ roughly coincides with $Kn = 1$. Therefore, the strongest effect is given at a pressure where the mean free path of the gas molecules equals the size of the particles. 
This view leads to a more natural division in low- and high-pressure region. 
The right hand part of the denominator in eq. \ref{eq:Rohatschek} or $p_{max}/{p}$ dominates at a pressure much
lower than $p_{max}$ or $Kn >> 1$ and the two first terms in the denominator become negligible. The force is proportional to the pressure in that case. $D$, which is present in the nominator in $F_{max}$ and in the denominator in $p_{max}$, cancels out. If $\delta$ is assumed to be 0, eq. \ref{eq:Rohatschek} using eq. \ref{eq:RohatschekParameters} simplifies to  }

\begin{equation}
	F_{Ph}=\frac{\pi a^3 I \alpha p}{6kT}.
	\label{eq:Photophorese}
\end{equation}

The size dependence $a^3$ in eq. \ref{eq:Photophorese} has two components. $a^2$ is showing the proportionality to the cross-section {for a spherical particle. This factor can be understood at low pressure or large Knudsen numbers (fig. \ref{fig:idea}a) as only the individual impacting gas molecule counts whose number increases linearly with cross-section; or $a^2$ for a spherical particle. 
Another $a$-dependence in eq. \ref{eq:Photophorese} comes from the temperature difference that develops from the illumination between front and back side. The thicker the particle, the larger the temperature difference, and the
larger the difference in momentum transferred by gas molecules at the different cold and warm sides. To a first order the
temperature difference and therefore the force is proportional to the thickness or $a$ for a spherical particle.} 

At high pressure net motion of gas molecules in the vicinity of a particle surface occurs. In general there is a net flow of gas from the cold side to the warm side of a non-equilibrated paticle along the surface {(see fig. \ref{fig:idea}b). This is called thermal creep and also leads to a force on a particle. The fundamental physics for this force is challenging and has been debated. However, it was shown recently that the gas density builds up on the warm side due to this gas flow and pushes the particle toward the cold side \citep{ketsdever2012} (\ref{fig:idea}b). }
{This high-pressure region of the photophoretic force can also be approximated with respect to the ambient pressure dependence. The first term in the denominator in eq. \ref{eq:Rohatschek} or ${p}/p_{max}$ dominates at a pressure much
higher than $p_{max}$, the other two terms are negligible and the force decreases with pressure as $1/p$ in that case.} {The decrease with ambient pressure arises because the thickness of the layer along the surface for which thermal creep occurs is on the order of the mean free path. This constantly diminishes at higher pressure. A back flow close to this layer can equilibrate the system more easily at high pressure (fig. \ref{fig:idea}c).} This already leads to the second group of particles, which have been studied over the past 100 years, the plates \citep{ketsdever2012}. For plates a similar pressure dependence as in eq. \ref{eq:Rohatschek}
has been suggested, that is, in the free molecular flow regime the pressure enters linearly ($\sim p$) \citep{ketsdever2012}. In the hydrodynamic regime it enters as $\sim 1/p$. However, for plates the maximum force pressure is not straightforward to define because it is 
$a$ $ priori$ unclear which size enters, or in other words the question might be which
value to use for the Knudsen number marking the transition. 

The plate diameter is relatively large compared with the thickness, and
choosing one value or the other will change the Knudsen number by one or more orders of magnitude. The situation even deteriorates for a plate with channels that introduce another size, namely the diameter of the channel.

Experiments dedicated to the pressure dependence of radiative forces on plates have  been carried out by \citet{sel2009}. The authors measured a transition (maximum force) at 
4 to 6$\times10^{-3}$\,mbar for plates heated not by light, but electrically. Two circular plates with diameters of 8.6 and 11.1\,cm, and a rectangular plate with the same surface area as the bigger disk with dimensions of 7.62\,\,$\times$\,12.7\,cm$^2$ were studied, all with a thickness of 0.95\,cm. 
As seen below, the ambient force maxima pressure  found by these authores fits
the characteristic size defined by our experiments here.

{The photophoretic force is low at very low as well as at very high
pressure. However, the important point is that it has a maximum at an ambient pressure related to typical size scales of a particle. If there are several 
prominent size scales that
vary by orders of magnitude, photophoretic forces will have different maxima. A significant force can still be related to an inner size scale (pore) at a given pressure even if
the overall size of an aggregate would suggest only a small contribution.}

\section{Experiments}

We carried out experiments similar to those of \citet{sel2009}, but heated the samples by light and extended the studies to plates with channels.

\subsection{Setup}

\begin{figure*}[tb]
	\centering
	\includegraphics[width=1.0\textwidth]{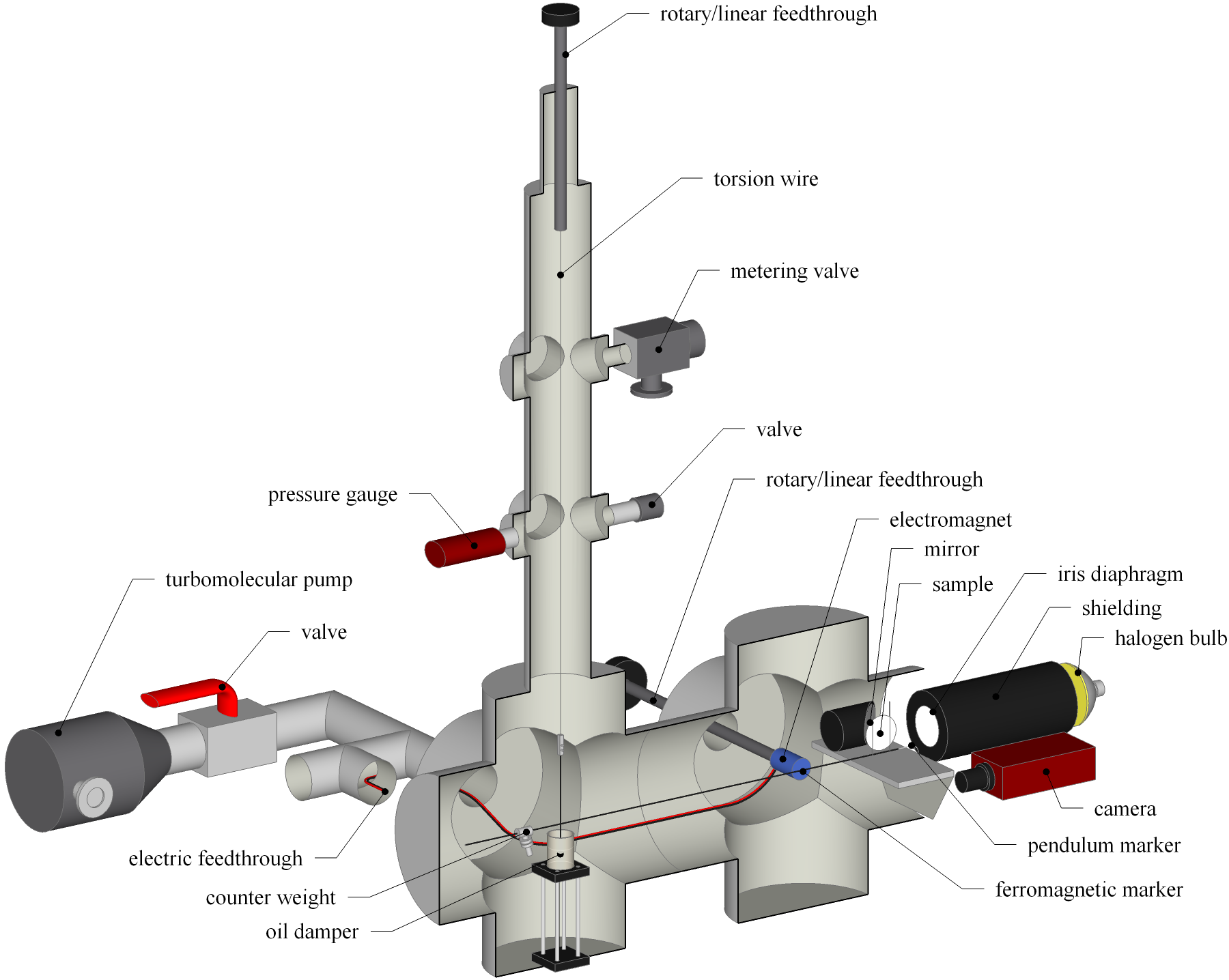}
	\caption{Sketch of the pendulum setup}
	\label{fig:setup}
\end{figure*}

To measure the photophoretic forces we developed a torsion balance inside a vacuum chamber that could be evacuated down to $1\times10^{-3}$\,mbar. The pendulum was fixed with an electric magnet and when switched off, a camera recorded its movement. A marker at the end of the pendulum was observed to trace its motion. Image analysis and averging over the marker pixels gives the deflection of the pendulum at a given time to a precision of $5\times{10^{-4}}\,{}^{\circ}$.
Knowing the elastic modulus of the torsion wire and the moment of inertia, we can calculate the force acting on the sample by measuring the acceleration of the pendulum. We determined the parameters of the pendulum prior to the experiments. With this setup we reached a resolution of 100\,nN.
A sketch of the setup is given in fig. \ref{fig:setup}. 

The sample was illuminated by a halogen lamp. The light was shaped by a single lens to achieve a uniform parallel light flux across the plates. The flux density at the position of the sample was measured to about $100\,\rm W/m^2$.
To adjust the pressure we used a metering valve, beginning at the lowest pressure and increasing the pressure inside the camber.

\subsection{Samples}

\subsubsection{Tissue plates}

\begin{figure}[tb]
	\centering
	\includegraphics[width=0.5\columnwidth]{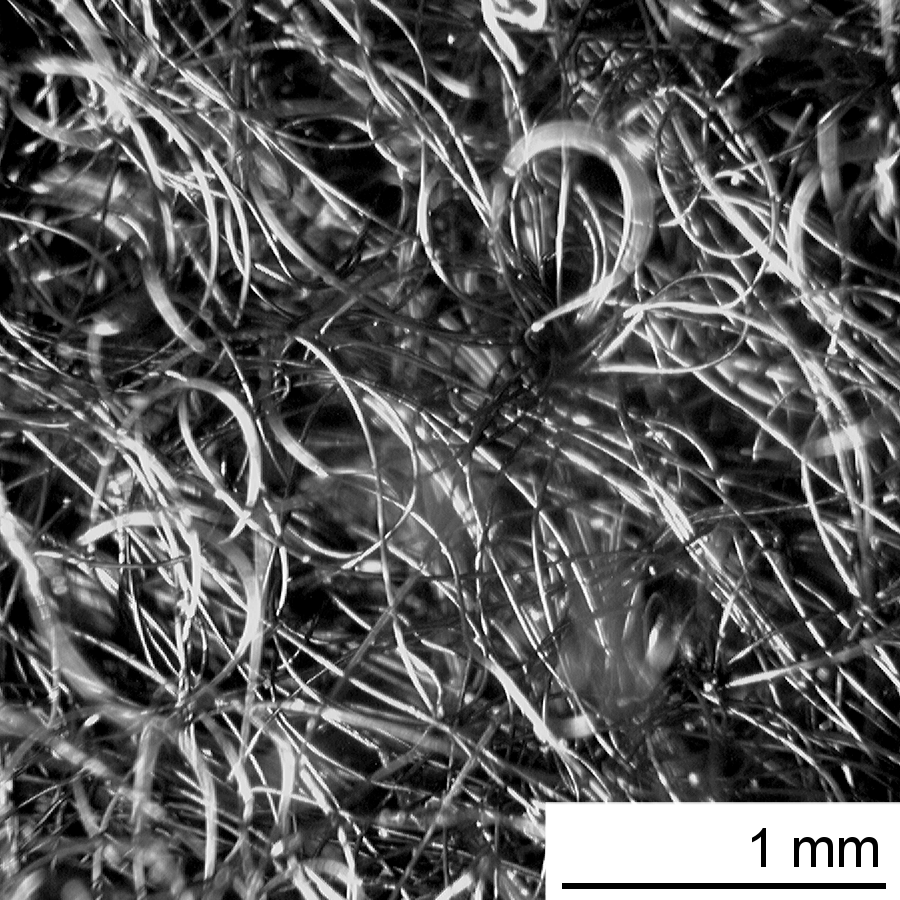}
	\caption{Microscope image of a tissue plate.}
	\label{fig:TissueMicroscope}
\end{figure}

As seen from eq. (\ref{eq:RohatschekParameters}) and (\ref{eq:Photophorese}) the photophoretic force depends on the thermal conductivity of the particle or plate.  In general, porous materials have low thermal conductivity. Tissue plates were chosen as one sample because they are easy to handle but still have a low thermal conductivity due to their porosity. The circular plates had radii of 10, 30, and 50\,mm. A microscopic view ot the tissue samples is given in fig. \ref{fig:TissueMicroscope}.
The individual fibres can be seen at microscopic resolution . The samples were also initially chosen in view of a potential effect of these pores on the photophoretic force. 
The porosity might influence the gas flow (i.e., act like channels) but the fibers of 
the tissue are so irregular and irregularly spaced that we did not find a well-defined pressure dependence due to the pores here, but only focusd on the effect of the total plate size of the tissue.

\subsection{Circuit boards}

\begin{figure}[htb]
	\includegraphics[width=\columnwidth]{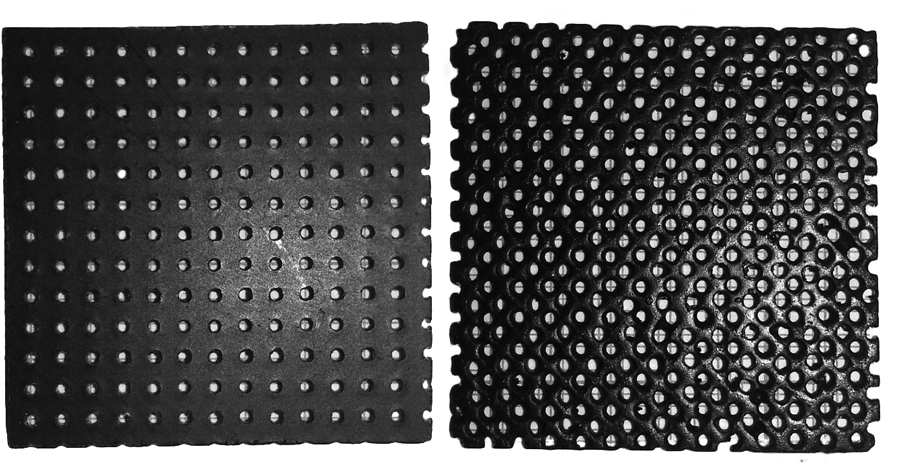}
	\caption{Samples made of circuit board. Their dimensions are 35\,$\times$\,35\,$\times$\,1.5\,mm$^3$.}
	\label{fig:CircuitBoardSamples}
\end{figure}

We then used two rectangular samples of circuit board with a grid of well-defined channels but reasonably low thermal conductivity. The two samples had dimensions of 35\,$\times$\,35\,mm$^2$ and a thickness of 1.5\,mm. The plates had 169 and 352 channels of 1 mm diameter to study the effect of well-defined channels on the photophoretic force. The plates were painted black to achieve a uniform light absorption. They are shown in fig. \ref{fig:CircuitBoardSamples}.

\section{Results}

\subsection{Tissue plates}

\begin{figure}[htb]
	\includegraphics{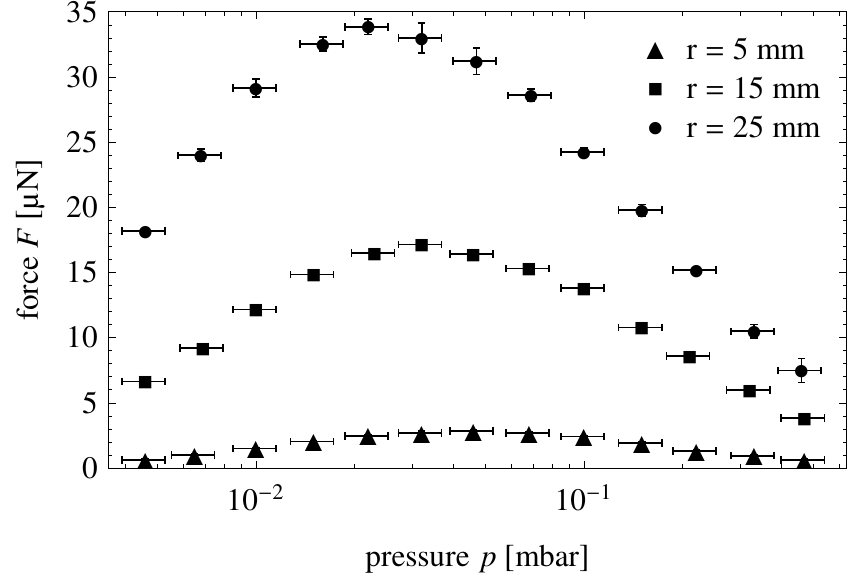}
	\caption{Pressure dependence of the photophoretic force on the tissue plates.}
	\label{fig:resulttissue}
\end{figure}

The pressure-dependent force on the tissue plates is plotted in fig. \ref{fig:resulttissue}.

\begin{figure}[htb]
	\includegraphics{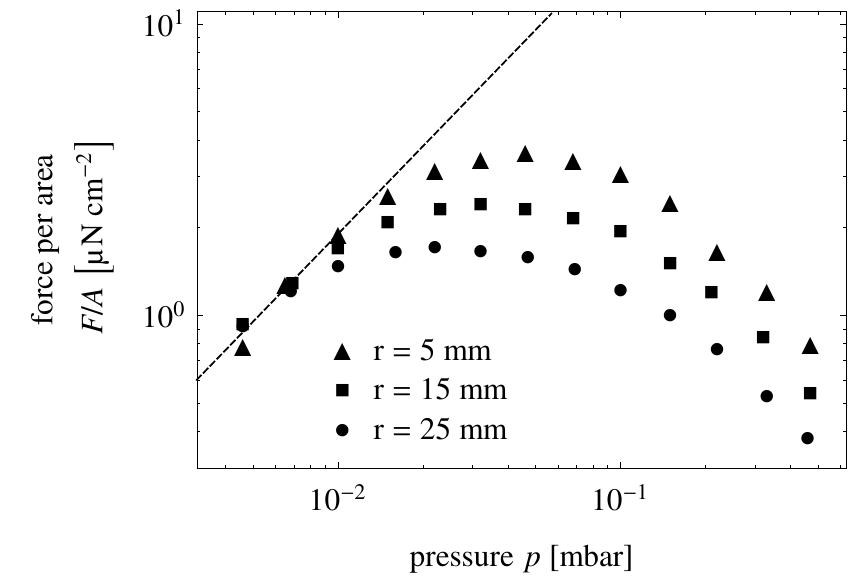}
	\caption{Pressure dependence of the photophoretic force per area on the tissue plates. The dotted line shows the linear dependency for low pressure}
	\label{fig:arearatio}
\end{figure}

\citet{sel2009} discussed the pressure dependence of the photophoretic force in terms of force per area or
force per edge. The photophoretic force per area is depicted in fig. \ref{fig:arearatio}. 
At low pressure, as expected and in agreement with \citet{sel2009}, we found a linear increase
with pressure of a constant force per area. 

\begin{figure}[htb]
	\includegraphics{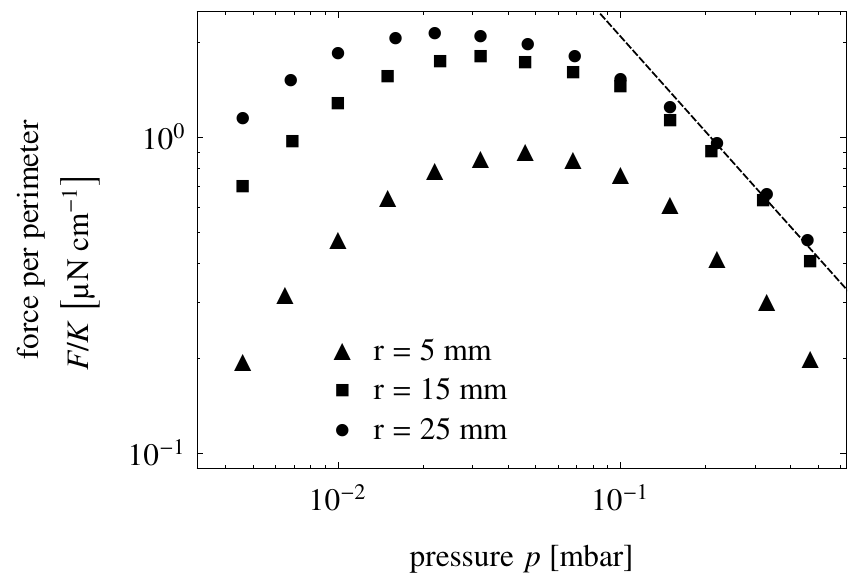}
	\caption{Pressure dependence of the photophoretic force per perimeter on the tissue plates; The line shows a linear dependency at high pressure for the two large samples converging.}
	\label{fig:rimratio}
\end{figure}

The photophoretic force per perimeter is shown in fig. \ref{fig:rimratio}. The two
largest samples converge at high pressure with a force per perimeter now decreasing
linearly with pressure (dotted line). This also agrees with the results by \citet{sel2009}. 
While the third plate converged well with the other two plates at low pressure for the
force per area, it is somewhat shifted to lower values here.
It has to be noted that this was the smallest plate. Because of its nature the tissue is somewhat compressible, which might be a reason for the shift. We therefore attribute the deviation to the specific sample. This is probably no general trend with plate size, but we cannot specify this in more 
detail at this point.

\begin{figure}[htb]
	\includegraphics{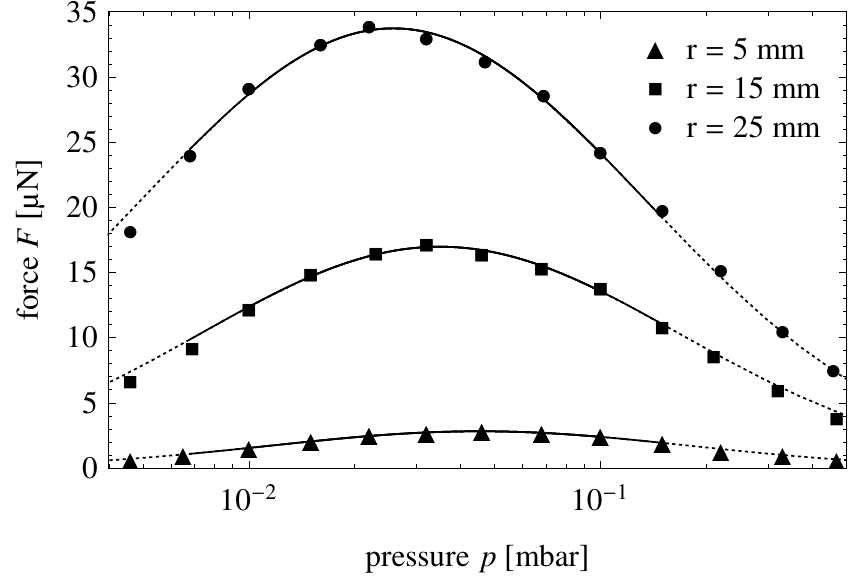}
	\caption{Pressure dependence fitted by eq. \ref{eq:Rohatschek} (dotted line). Only data points with at least two thirds of the peak value were considered for the fit (solid line) to reduced the chance of including other effects (convection, etc.).}
	\label{fig:tissuemax}
\end{figure}

For a spherical particle the photophoretic force depends on the 
pressure according to eq. 	\ref{eq:Rohatschek}. 
Here, $p_{max}$ for a spherical particle depends inversely on the radius of the sphere. To test if a plate can be described by
a typical size as well, we used eq. \ref{eq:Rohatschek} and determined a pressure 
$p_{max}$ of the maximum force (fig. \ref{fig:tissuemax}).
Despite the two different size scales present in a plate (diameter and thickness), the fit describes the
transition regime very well.

\subsection{Circuit boards}

\begin{figure}[bth]
	\centering
	\includegraphics{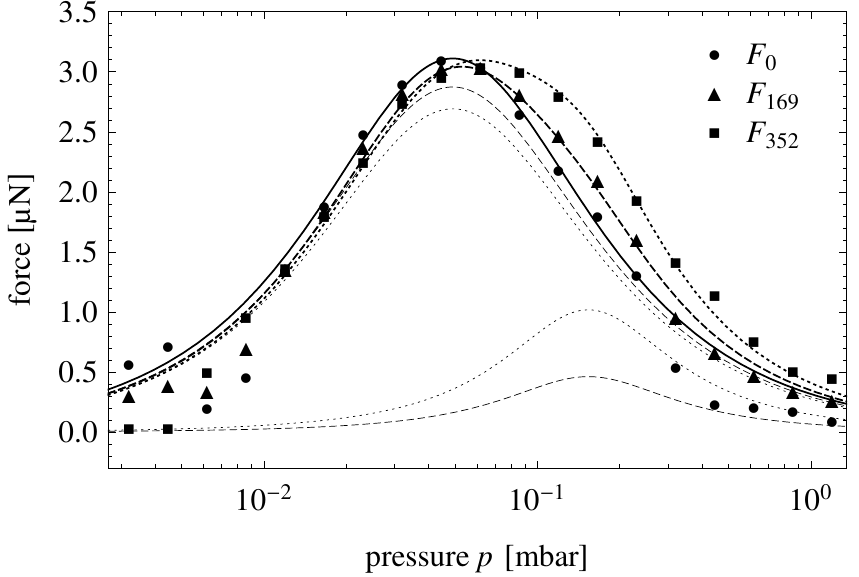}
	\caption{Forces measured for channel plates ($F_{169}$ and $F_{352}$) and extrapolated to a channel-less plate ($F_0$). 
	The thick lines are the sum of the associated thin lines of the two component fits. The thick solid line is the fit of the extrapolated force for a plate without channels.} 
	\label{fig:HolesFit}
\end{figure}

The data on the circuit boards are displayed in fig. \ref{fig:HolesFit}. At pressure
higher than $10^{-1}$ mbar the force on the plate with the larger number of 
channels is significantly increased.  To separate the contributions from the whole plate and the channels, we took the following
approach: 

We first calculated the difference between the two channel plates.
We attributed this difference (increase) to the presence
of 183 more channels in the second plate. 
We considered that a whole plate (not available for measurements) would also differ by the 
contributions of 169 channels to the plate with 169 channels. Therefore, we constructed an artificial dataset for a whole plate without channels,
subtracting the difference again (in detail 169/183 of it) from the dataset of the 169-channel plate.
These data were fitted by eq. \ref{eq:Rohatschek} in analogy to the tissue plates. This 
whole procedure was only carried out to obtain the best value for the pressure at the force maximum due to the total plate size. We then assumed the resulting pressure to be fixed to describe the pressure dependence on the total plate size for the two channel plates. 

We added a second function according to eq. \ref{eq:Rohatschek} to model the contribution of the channels.
Because the channels in the two plates have the same dimensions and only differ in their number, the position of the pressure for the maximum force 
should be the same in both cases and only the absolute value of the force should vary. With these assumptions we gained the overall fits plotted in fig. \ref{fig:HolesFit}. The channels give an additional force at a different pressure. 

The magnitude of the two contributions can be compared with the plate area and the perimeter length of the according samples, which is listed in table \ref{tab:ratios}. For the total cross-section the channel cross-sections are subtracted from the total plate area. For the perimeter every channel perimeter is added to
the outer perimeter of the plate.

\begin{table}[hbt]
	\centering
	\begin{tabular}{l@{\hspace{1pt}}rr@{\hspace{2pt}}rr@{\hspace{2pt}}r}
		\hline \hline
		\rule{0pt}{9pt} & no channels & \multicolumn{2}{c}{169 channels} & \multicolumn{2}{c}{352 channels} \\
		\hline
		\rule{0pt}{9pt}area [cm$^2$] & 12.25 (100\%)& 11.27 & (92.0\%) & 10.17 & (83.0\%) \\
		\rule{0pt}{9pt}$F_{max,P}$ [$\mathrm{\mu N}$] & 3.11 (100\%)& 2.87 & (92.3\%) & 2.69 & (86.5\%) \\
		\hline
		\rule{0pt}{9pt}perimeter [cm] && 62.3 & (54.2\%) & 115.0 & (100\%) \\
		\hline
		\rule{0pt}{9pt}channels & &169 & (48.0\%) & 352 & (100\%) \\
		\rule{0pt}{9pt}$F_{max,C}$ [$\mathrm{\mu N}$] && 0.461 & (45.3\%) & 1.02 & (100\%) \\
		\hline \hline
		
	\end{tabular}
	\caption{Relations between force maxima, cross sections and perimeters}
	\label{tab:ratios}
\end{table}

The low-pressure maximum $F_{max,P}$ scales with the plate cross-section. Here, the only influence of the channels is to decrease the total cross-section. The total perimeter is not important in the low-pressure data. The magnitude of the force attributed to the channels $F_{max,C}$ scales linearly with the number of channels. The higher the number of channels, the stronger the force.

\subsection{Pressure of maximum force}

The data give different values of the pressure at which the force is at maximum.
The position of the maximum force differs stongly with changing dimensions of the samples.
We estimate here if a potential systematic description of the maximum force pressure is possible. The following data points of different sized samples can be considered:

\begin{itemize}
\item cylindrical tissue plates
\item square circuit boards
\item cylindrical channels within the circuit boards
\item data on square plates reported by \citet{sel2009}
\item two data points from glass plates consisting of sintered microspheres studied in microgravity experiments by \citet{Kuepper2013}.

\end{itemize}

We began with the 
most simple relation known for spherical particles, given as $p_{max} \sim a^{-1}$, where $a$ is the particle radius \citep{rohatschek1995}. Because non-spherical particles do not have a single dimension, but the fits describing them are consistent with the description based on a single parameter, i.e. $p_{max}$, there should be one characteristic length $l$, so that $p_{max} \sim l^{-1}$.

\begin{figure}[htb]
	\centering
	\includegraphics{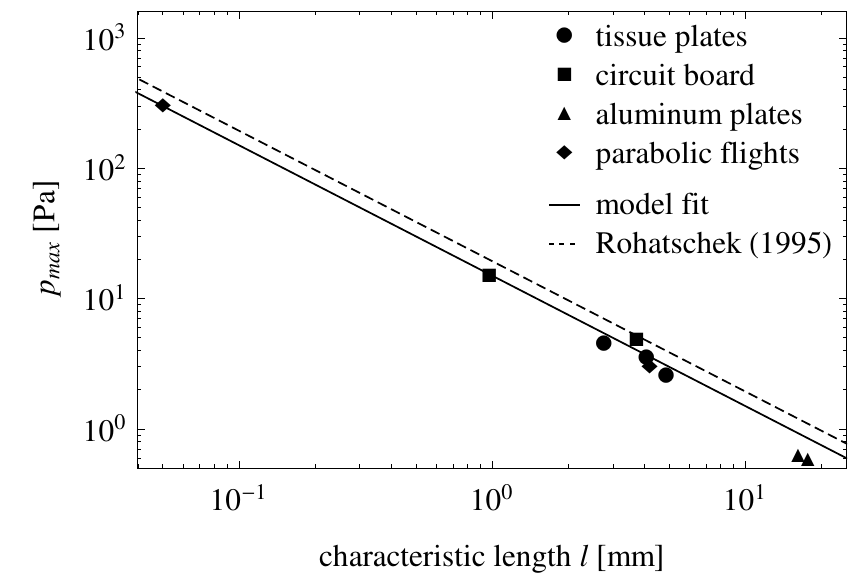}
	\caption{Maximum forces pressure for various samples with respect to the characteristic length. The solid line is a fit of the form
	$p_{max} = 1/l$. The dashed line shows the calculated position of the maximum force by eq. (\ref{eq:RohatschekParameters}). triangles: data taken from \citet{sel2009}; diamonds: data taken from \citet{Kuepper2013}}
	\label{fig:PositionOfMaximumForce}
\end{figure}

For the cylindrical plates there are two dimensions to consider, radius and thickness. To maintain the overall dimension of a length and obtain a comparable quantity we used a weighted geometric mean of radius $r$ and thickness $d$ with the form
\begin{equation}
	l=r^{x}\cdot d^{1-x},
	\label{eq:geometricMean}
\end{equation}
where $0<x<1$ is a parameter to deduce from the experimental data.
The plates are rectangular, but \citet{sel2009} found that the force on circular plates and rectangular plates with the same area are subject to the same force, therefore we used the radius of a circle with the same area. For the size for the plates consisting of sintered glass spheres by \citet{Kuepper2013} we used the total plate radius. For the channel size of the sintered plate we used the radius of the constituent spheres, which
approximately equals the typcial size of the pores. This is not quite as well defined in
dimensions as the channels in our experiments. The
fitted parameters did not change significantly, when we left these data out.
In total, a value of $x=0.35$ was obtained by a least-squares fit,
as depicted in fig.	\ref{fig:PositionOfMaximumForce}. 

For the given experimental data we derive the fit
\begin{equation}
	p_{max}=\frac{15\,\mathrm{Pa\,mm}}{r^{0.35}\cdot d^{0.65}}.
\end{equation}
From eq. \ref{eq:RohatschekParameters} we can obtain the position of the maximum force for spherical particles, and using the calculated characteristic length, we can compare this with our measurements. For these calculations we used $T=296$\,K, $R=8.3\,\rm J/(mol\,K)$, $M=29\,\rm g/mol$ and $\eta=1.8\times 10^{-5}\,\rm Pa\,s$. In this case the calculated proportionality constant is 19.4\,Pa\,mm.

Obviously, the photophoretic forces on plates and channels can (at least formally)  be clearly quantified by one scheme for a characteristic size.
Especially for the channels considered as kind of inverse plates this was not necessarily to be expected.

\section{Motion of large porous bodies in protoplanetary disks}

The experiments show that in the simplest version the photophoretic force on large and highly porous bodies is a superposition of two photophoretic forces:

\begin{itemize}
\item a force related to the total size
\item a force related to the pores or building-block size
\end{itemize}

The important point is that both forces have the characteristic pressure dependence related to the relevant size and their strength peaks at different pressure, and, therefore, it peaks at different radial distances to the star. The scaling of both forces relative to each other, $f_s$, will depend on the details of the inner structure of the porous bodies.
These findings can be applied to protoplanetary disks if the force 
is written as the 
sum of two equations of the Rohatschek type (eq. \ref{eq:Rohatschek}). 
For the subsequent application we assumed the parameter $\delta$, which is small, to be $0$ and introduced the above-mentioned scale factor $f_s$ to describe the fraction of the force attributed to the pores related to the force attributed to the total size, or
\begin{equation}
	F_{Ph} = \frac{2 F_{large}}{\frac{p}{p_{large}} + \frac{p_{large}}{p}}
	+ f_s \cdot \frac{2 F_{small}}{\frac{p}{p_{small}} + \frac{p_{small}}{p}}.
	\label{eq:Rohsumme}
\end{equation}
Here, large refers to the overall size of the body. Small relates to the pore size or size of the building blocks. 
Other parameters can be taken from eq. \ref{eq:RohatschekParameters}.

To see the significance of each of the two components and the difference introduced
by porosity we considered the drift of 
large bodies in a protoplanetary disk caused by photophoretic force, gravity, and 
gas drag.
We mostly followed the work by \citet{Weidenschilling1977}. 

The difference of the transversal motion of the gas with respect to Keplerian 
speed can be considered in terms of a residual gravity for solids $\Delta g$
\begin{equation}
\Delta V = V_k-V_g\cong-\frac{\Delta g}{2g}V_k,
\end{equation}
where $V_k$ is the Kepler velocity and $V_g$ the gas velocity. 
For a disk where the density  (locally) decreases by a power law $p=p_0 r^{-n}$, the residual gravitational acceleration can be expressed as
\begin{equation}
\Delta g =-\frac{n R T}{\mu r},
\end{equation}
where
$\mu$ is the molecular weight of the gas and $R$ the gas constant. Note that $\Delta g<0$ means inward drift. When we add photophoresis, we obtain a radial acceleration 
\begin{equation}
\Delta a =-\frac{n R T}{\mu r}+\frac{F_{ph}}{m}.\\
\end{equation}
In equilibrium, this acceleration on a particle is balanced by the gas drag force $F_D$. The radial inward drift velocity $u$ and the transversal drift velocity $w$ with respect to the gas add to the total relative motion with respect to the gas of $v=\sqrt{u^2+w^2}$. Following \citet{Weidenschilling1977} this is
\begin{equation}
\frac{F_D}{m}\frac{u}{v}+\Delta a+ \frac{2 w}{r} V_k=0 \\
\end{equation}
and
\begin{equation}
\frac{F_D}{m}\frac{w}{v}-\frac{u V_k}{2 r}=0,
\label{eq:Weid2}
\end{equation}
 which yields for $u$
 \begin{equation}
u=\pm \sqrt{-4w^2-2\Delta a\frac{r}{V_k}w}.
\label{eq:uw}
\end{equation}
Inserting eq. \ref{eq:uw} into eq. \ref{eq:Weid2},
one obtains an equation for $w$. This  equation depends on the assumed drag law for $F_D$. The drag law itself depends on the solution of the equation. To solve this consistently, the solutions for all drag laws were calculated and the appropriate solution was chosen. Table \ref{tab:DragLaws} summarizes the used drag laws and the conditions for their validity. There are two modifications with respect to the aproach \citet{Weidenschilling1977} used. 

First, for some parameters, the drift velocities and Reynolds numbers
 become high. To smooth the transition between the Epstein to the Newtonian regime we calculated the point where both drag forces are equal and located the transition there. Second, for some parameters no self-consistent solution was obtained when the Reynolds number was only slightly larger than 1. This is also due to the choice of four drag forces, which do not connect smoothly. In this cases the Stokes regime was used even though the Reynolds number increased to 40. These adaptations do not change the basic results.
\begin{table}[hbt]
	\centering
	\begin{tabular}{lcr}
		\hline \hline
Drag regime &$F_D$ & Validity\\ \hline
Epstein & $\frac{4\pi}{3}\rho a^2 v \bar{c}$ & $\frac{\lambda}{a}>4/9$ \\
& & $a \le \frac{27 \sqrt[3]{3} \pi^{\frac{5}{6}} \eta \left( \frac{R T}{\mu}\right)^{\frac{1}{6}} v^{\frac{2}{3}}}{8 \sqrt[2] p}$ \\
High Reynolds number & $ 0.44 \pi a^2 \rho \frac{v^2}{2}$ & $\mathit{Re}\ge 800$\\
Newtonian flow & $24 \mathit{Re}^{-0.6} \pi a^2 \rho \frac{v^2}{2}$ & $1< \mathit{Re}<800$\\
Stokes & $24 \mathit{Re}^{-1} \pi a^2 \rho \frac{v^2}{2}$ & $\mathit{Re}<1 (40)$ \\
		\hline \hline
	\end{tabular}
	\caption{Drag laws}
	\label{tab:DragLaws}
\end{table}

\begin{figure}[htb]
	\centering
	\includegraphics[width=1.0\columnwidth]{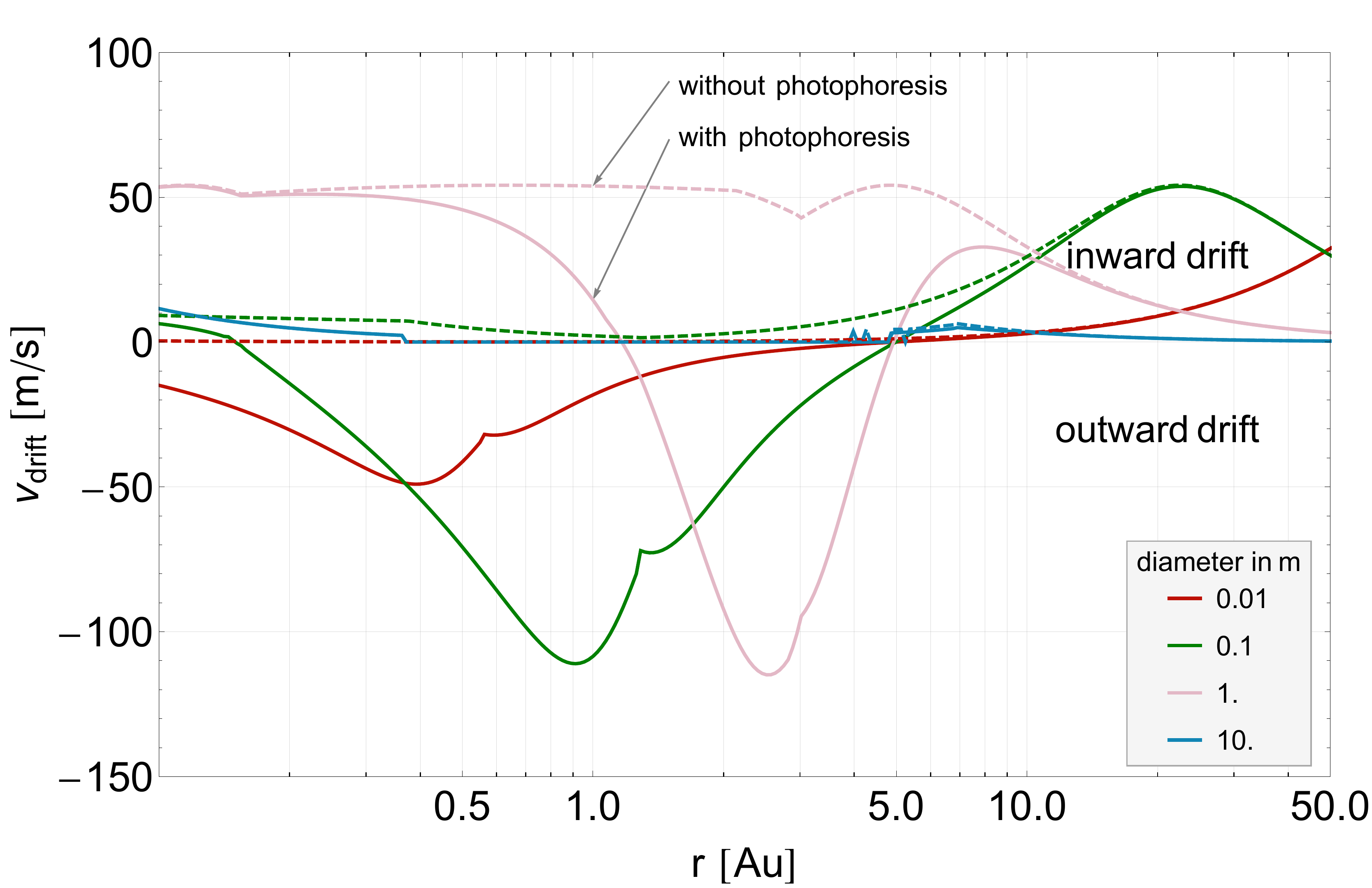}
	\caption{Drift of different sized \textbf{non porous} bodies in protoplanetary disks
	(minimum mass solar nebula).
Dashed line: without photophoresis, solid line: with photophoresis}
	\label{fig:pore1}
\end{figure}

\begin{figure}[htb]
	\centering
	\includegraphics[width=1.0\columnwidth]{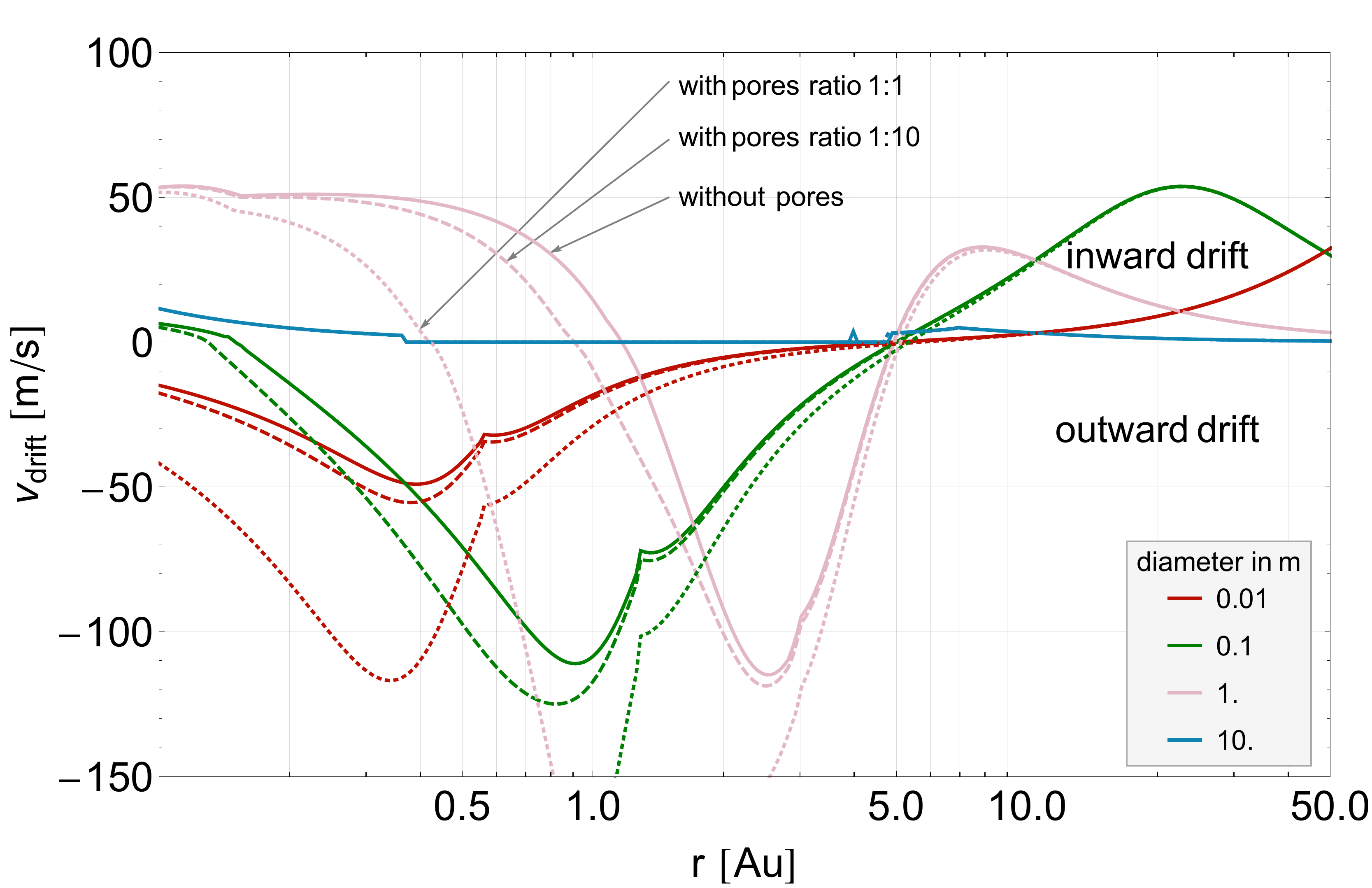}
	\caption{Drift of differently sized bodies including porosity.
Upper dashed line: non porous bodies including photophoresis (as fig. \ref{fig:pore1}), lower lines: with second photophoretic
term attributed to the pores. We assumed the photophoretic strength ratio $f_s$ to be 0.1 or 1}
	\label{fig:pore2}
\end{figure}

We used these equations to calculate the radial drift velocity of bodies of 
1cm, 1dm, 1m, and 10m. Without photophoresis and applying the \citet{Weidenschilling1977} nebula model we obtained the same results as \citet{Weidenschilling1977}, except that our two modifications smooth the 
transition regions between different drag regimes (not plotted here). Using the minimum mass solar nebula by \citet{hayashietal1985}, we derived fig. \ref{fig:pore1} and \ref{fig:pore2}. In this first application we fixed the small size scale at 1 mm radius, which in view of abundant mm-chondrules in primitive meteorites is at least
one reasonable choice and useful as an example to show the influence of the pore
size scale.
As can be seen,  introducing  porosity has a strong effect. At 1 AU, for instance, it can change the sign of the drift and move meter-sized bodies outward, which are otherwise often considered to problematic because of their large inward drift. This concept holds
for other sizes as well. A ten-meter-body, however, is not influenced much by photophoresis, regardless of wheather it has mm-pores.

We note that these calculations are only applicable in optically thin environments. This
might occur at later times, before disk dispersal, but probably the gas density is also lower then. The other active photophoretic site is close to the illuminated inner edge of protoplanetary disks, where photophoresis is always acting. In this case photophoresis and pores will lead to a sorting of bodies in the transition
zone between optically thin and optically thick outward parts of the disk. 
However, the calculations clearly show the strength of photophoresis itself and 
especially the effect of the pores, which for instance supports meter-sized bodies, which are assumed to drift fastest without photophoretic support.

\section{Conclusions}

With measurements of the force acting on plates with a torsion balance, we were able to quantify the photophoretic force and to distinguish the photophoretic force on the whole plate from an additional force caused by channels within the plate. The magnitudes of both contributions scale with
the areas of the whole plate and the area covered by the channels.

The position of the pressure at which the photophoretic force is at maximum depends on the geometric properties. For spherical particles
studied in the past it is proportional to $r^{-1}$ \citep{rohatschek1995}. We found that we can use the same simple inverse dependency, but attributing a characteristic length $l$ to the plates and channels. This length combines the radius of the plate (channel) and the thickness as 
$	p_{max} \sim r^{-0.35}\cdot d^{-0.65}$. The photophoretic force reaches its maximum roughly at the pressure for which the mean free path equals the characteristic length.

The measurements reported here are a first step toward modeling photophoretic transport of larger bodies in protoplanetary disks, that is at the inner edge of a disk. Because larger bodies are often ruble piles consisting of smaller units, the strength of the photophoretic force will depend on the 
structure of these bodies. Application to the minimum mass solar nebula showed that porosity, ideally simulated by channel plates in our experiments, 
can lead to a significant increase in the photophoretic force at low pressure even
if the total size of the body would suggest the photophoretic force to decrease.
An instructive example is that motion of meter-sized bodies -- assumed to drift inward --
can be reversed by photophoretic motion, but only if porosity is taken into account.
\begin{acknowledgements}
This work is based on DFG funded projects within the research group FOR 759 and the priority program SPP 1385. Access to parabolic flights, providing data here, has been granted by ESA education within the "Fly your thesis program 2012". We also thank the referee for the thoughtful review of this paper.      
\end{acknowledgements}

\bibliographystyle{aa} 

\end{document}